\documentclass[twocolumn,showpacs,preprintnumbers,amsmath,amssymb]{revtex4-1}
\usepackage{graphicx}
\usepackage{dcolumn}
\usepackage{bm}

\begin{document}


\title{Stripes and honeycomb lattice of quantized vortices in rotating two-component Bose-Einstein condensates}

\author{Kenichi Kasamatsu and Kouhei Sakashita}
\affiliation{
Department of Physics, Kindai University, Higashi-Osaka, Osaka 577-8502, Japan
}

\date{\today}

\begin{abstract}
We study numerically the structure of a vortex lattice in two-component Bose-Einstein condensates with 
equal atomic masses and 
equal intra- and inter-component coupling strengths. The numerical simulations of the Gross-Pitaevskii equation 
show that the quantized vortices form uncertain lattice configurations accompanying the 
vortex stripes, honeycomb lattices, and their complexes. This is a result of the 
degeneracy of the system for the SU(2) symmetric operation, which makes 
a continuous transformation between the above structures. 
In terms of the pseudospin representation, the complex lattice structures are identified to 
a hexagonal lattice of doubly-winding half-skyrmions. 

\end{abstract}


\maketitle

\section{Introduction} \label{intro}
Quantized vortices are the basic constituents of superfluid hydrodynamics, 
having a definite quantized circulation. 
When a superfluid is subject to an external rotation, the superfluid 
forms a lattice of quantized vortices, creating rigid-body rotation. 
In usual superfluids with the scaler order parameter, the vortices form a triangular Abrikosov lattice \cite{Sonin}. 
Conversely, for unusual superfluids characterized by, e.g., 
multiple order parameters, a rich variety of 
vortex lattice structures can emerge, because the vortices in such superfluids have an 
complex core structure and there are multiple scales of 
interactions between them \cite{Tunerev}. 

Multicomponent superfluids have been realized by cold atomic 
Bose-Einstein condensates (BECs) \cite{Pethickbook}, and the vortex structures have been 
studied very well \cite{Kasarev2}. 
The observation of vortex lattices in rotating multicomponent 
BECs has been reported in Ref. \cite{Schweikhard}. 
It has been known that rotating two-component BECs have a rich vortex lattice structure, which is particularly dependent on the intercomponent interaction 
\cite{Mueller,Kasamatsu,Barnett,Mason,Hsueh,Wei,Aftalion,Kuopanportti,Cipriani,Ghazanfari,Kumar,Uranga}. 
In cold atom experiments, the intercomponent interaction may be considered as a tunable parameter \cite{Thalhammer,Papp}.
In the absence of intercomponent interaction, the two condensates 
behave independently and the vortex lattice then forms a triangular 
lattice, as in conventional scalar condensates. 
When the intercomponent coupling is positively increased, the repulsive interaction 
between two components displaces the lattice locations, thereby decreasing 
the overlapping of the condensate densities. 
Then, the triangular lattice deforms into an interlaced square lattice. 
A further increase of intercomponent repulsion induces the phase 
separation of the two components.
Subsequently, the periodic structure of the vortex lattice transforms into 
interwoven vortex sheets \cite{Kasamatsu,Kasashet}. 
Alternatively, one can consider the attractive force between 
two components by decreasing the intercomponent coupling to a negative value. 
Then, the location of the vortices is locked to the same position \cite{Mueller,Barnett,Kuopanportti}. 

It is noticeable that, when the all intercomponent and intracomponent couplings 
are equivalent, with the hamiltonian having an exact SU(2) symmetry, one peculiar structure 
of the vortex lattice appears, namely, honeycomb and double-core lattices. 
This has been observed in numerical simulations of the coupled 
Gross-Pitaevskii (GP) equations \cite{Kasamatsu,Hsueh} and Monte-Carlo simulation 
of similar models \cite{Galteland}. 
Conversely, the theoretical analysis based on the lowest Landau level approximation has predicted that the vortex stripe, 
the alternating rows of vortices in each component, is the stable structure \cite{Mueller}.
So far, there is no theoretical interpretation of the problem of what are
true stable structures in this situation. 
In this paper, we demonstrate that both the stripe and the honeycomb lattice are true stable structures of fast rotating 
two-component BECs with an SU(2) symmetry. We find that these structures are connected through the 
global SU(2) rotation and identical to a lattice of doubly-winding half-skyrmions in the pseudospin picture. 

This paper is organized as follows. After introducing the theoretical formulation in 
Sec.~\ref{formulation}, we first provide some numerical evidence of the vortex 
lattice structures in Sec.~\ref{numerical}. Following this numerical observation, 
we attempt to explain the mechanism of why the honeycomb lattice structure 
appears in two-component BECs. 

\section{Vortex states in rotating two-component BECs}

After introducing the theoretical formulation for describing the two-component BECs, we provide a brief account of the vortex lattice structures in this system through a numerical simulation of the GP equations. 

\subsection{Coupled Gross-Pitaevskii equations} \label{formulation}
The equilibrium solutions of vortex states in rotating two-component BECs can be obtained by a minimization 
of the GP energy functional 
\begin{align}
E[\Psi_1,\Psi_2] = \int d \bm{r} \sum_{i=1,2} \Psi_i^{\ast} \left( \hat{h}_i - \Omega \hat{L}_z \right) \Psi_i + E_\text{int}\\ 
E_\text{int} =\int d \bm{r} \left( \frac{g_1}{2} |\Psi_1|^4 + \frac{g_2}{2} |\Psi_2|^4 + g_{12} |\Psi_1|^2 |\Psi_2|^2 \right) \label{interactionene}
\end{align}
in a rotating frame with a rotation frequency of $\mathbf{\Omega}=\Omega \hat{\bm{z}}$. 
Here, $\hat{h}_i =  - \hbar^2 \nabla^2/(2m_i)  + V^i_\text{ext}(\bm{r})$ is a single-particle hamiltonian, 
$m_i$ is the atomic mass of the $i$-th component ($i=1,2$), and the interaction 
strengths are given as $g_{i} = 4 \pi \hbar^2 a_{i}/m_i$ and $g_{12} = 2 \pi \hbar^2 a_{12}/m_{12}$ 
with the intra- and inter-component $s$-wave scattering lengths $a_i$ and $a_{12}$ and the 
reduced mass $m_{12}^{-1}=m_{1}^{-1} + m_{2}^{-1}$.
To discuss the lattice structure, it is enough to confine ourselves to 
analyzing the equation in a two-dimensional (2D) $x$-$y$ plane. 
To scale the equation, we introduce the length and time scales of the trapping 
potential $V_{\rm ext}^{i} = m_{i} \omega_{i}^{2} r^{2}/2$ as 
$a_{\rm ho}=\sqrt{\hbar/2 m_{12} \bar{\omega}}$  
and $\bar{\omega}^{-1}$, respectively, with $\bar{\omega}=(\omega_{1}+\omega_{2})/2$. 
In a 2D system, the wave function is normalized by the particle number $N_{i}^{\rm 2D}$ ($i=1,2$) 
in 2D as $\Psi_{i} \rightarrow \sqrt{N_{i}^{\rm 2D}}\Psi_{i}/a_{\rm ho}$. 
We then obtain the dimensionless GP equation: 
\begin{align}
\biggl[ - \frac{m_{12}}{m_{1}} \nabla^{2} + \tilde{V}_{1} + C_{11}|\Psi_{1}|^{2} + C_{12}|\Psi_{2}|^{2} - \tilde{\Omega} L_{z} \biggr] \Psi_{1} = \mu_{1} \Psi_{1}, \label{bindimless1} \\
\biggl[ - \frac{m_{12}}{m_{2}} \nabla^{2} + \tilde{V}_{2} + C_{22}|\Psi_{2}|^{2} + C_{21}|\Psi_{1}|^{2}  - \tilde{\Omega} L_{z}\biggr] \Psi_{2} = \mu_{2} \Psi_{2}. \label{bindimless2} 
\end{align}
Here, the rotation frequency is $\tilde{\Omega}=\Omega/\bar{\omega}$, 
and the trapping potential is $\tilde{V}_{i} = \frac{1}{4} \frac{m_{i}}{m_{12}} \frac{\omega_{i}^{2}}{\bar{\omega}^{2}} (x^{2}+y^{2})$.
The interatomic interactions for the intra- and inter-component are 
written as: 
\begin{equation}
C_{ii} = 8 \pi \frac{m_{12}}{m_{i}} N_{i}^{\rm 2D} a_{i} , \hspace{4mm} C_{ij} = 4 \pi N_{j}^{\rm 2D} a_{12}  \hspace{2mm} (i \neq j),
\end{equation}
where the s-wave scattering length is confined to be positive. 
Given that the particle number of each component is conserved, the chemical 
potential $\mu_i$ is determined by a normalization of the wave function $\int dx dy |\Psi_{i}|^{2} = 1$. 

\subsection{Vortex lattice phase: numerical study}\label{numerical}
Here, we briefly mention the properties of vortex lattices in rotating two-component 
BECs. In this study, our interest is focused on the vortex states when the hamiltonian 
has an exact SU(2) symmetry. Thus, we confine ourselves to the parameters 
$m_1 = m_2 =m$, $\omega_1 = \omega_2 = \omega$, $N_1^\text{2D}=N_2^\text{2D}=N$, 
and $C_{11} = C_{22} = C > 0$. Fixing $C=4000$, we search the equilibrium solutions of 
Eqs.~\eqref{bindimless1} and \eqref{bindimless2} by changing $\tilde{\Omega}$ 
as well as $\delta = C_{12}/C$ in the vicinity of unity. 

Even for $\tilde{\Omega} = 0$ the equilibrium solutions of Eqs.~\eqref{bindimless1} and 
\eqref{bindimless2} exhibit a rich variety of structures, depending on the various 
parameters of the system \cite{Kasarev2}. A salient feature is the occurrence of phase separation. 
In our parameter setting, the two components are miscible for $\delta<1$ and immiscible  
for $\delta > 1$. 

The properties of the vortex phases are also different for these two situations. 
In the miscible regime, the vortices form an interlaced triangular or 
square lattice depending on the ratio of the coupling strengths $\delta$ and the 
rotation frequency $\tilde{\Omega}$. The triangular lattice is a conventional lattice structure 
seen in rotating superfluids and type-II superconductors under a magnetic field, 
but the square lattice is an exotic structure in multicomponent superfluids. 
The transition of the lattice structure in two-component BECs has been discussed 
theoretically by using variational analysis based on the lowest Landau level 
expansion \cite{Mueller} and the argument based on the vortex-vortex interaction \cite{Aftalion}. 
In the immiscible regime, the phase separation 
favors the formation of vortex sheets or rotating droplets when the condensates 
are subject to external rotation \cite{Kasashet}. 
These behaviors can be understood through the ferromagnetic or antiferromagnetic nature 
of the interactions between coreless vortices \cite{Kasarev2}. 
Our focus is the vortex lattice structure at the boundary between 
the miscible and immiscible regimes. This situation is approximately realized 
in the experiments of two-component BECs with $^{87}$Rb atoms \cite{Hall}.

\begin{figure}[ht]
\centering
\includegraphics[width=1.0\linewidth, bb=0 0 482 652]{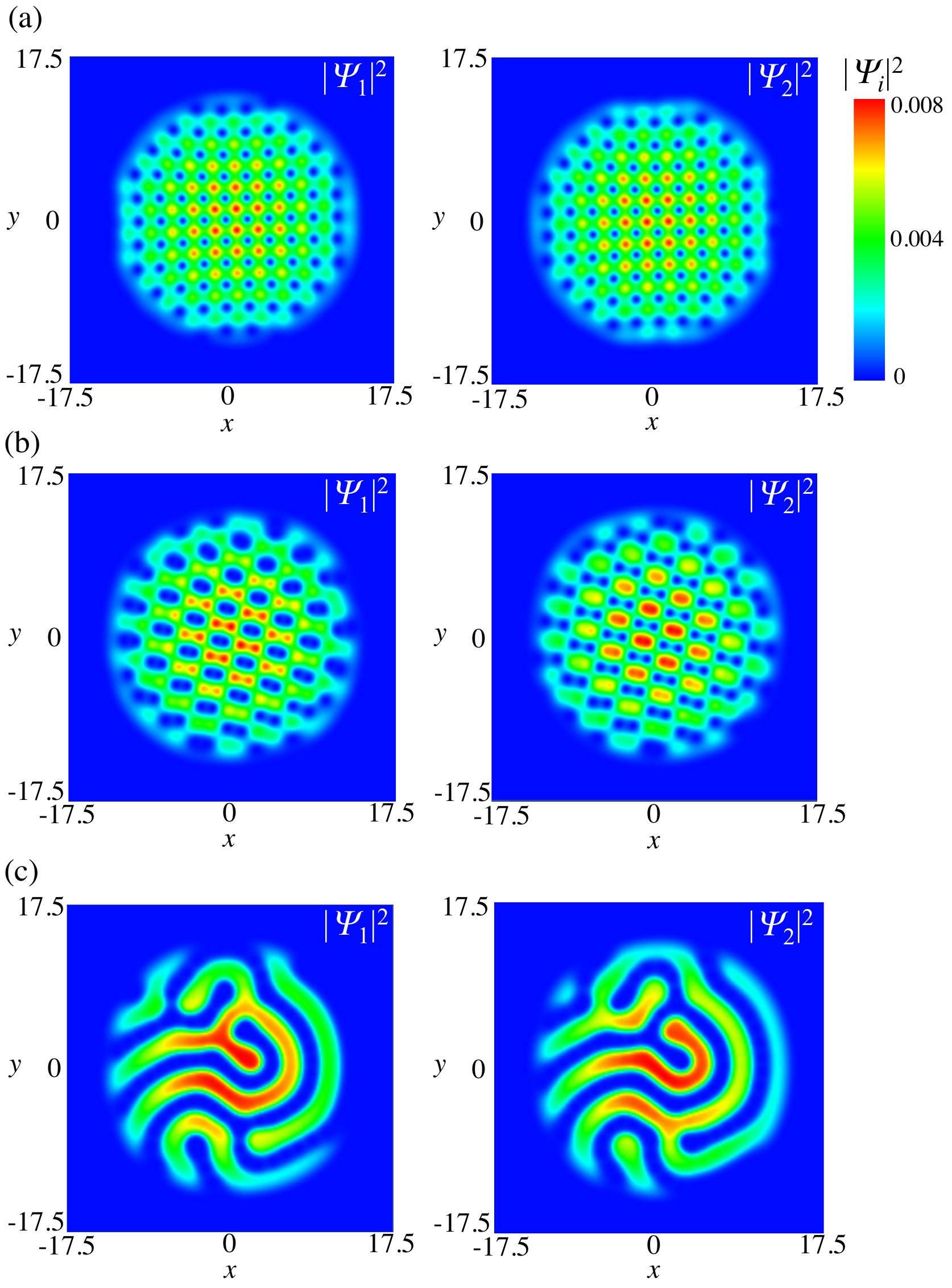}
\caption{(Color online) The density profile $|\Psi_1|^2$ (left) and $|\Psi_2|^2$ (right) of the 
equilibrium solutions of Eqs.~\eqref{bindimless1} and \eqref{bindimless2} 
for $C=4000$, $\tilde{\Omega}=0.8$, and (a) $\delta = 0.9$, (b) $\delta = 1.0$, and (c) $\delta=1.1$.}
\label{saka1}
\end{figure}
We first show the numerical results of the vortex lattice in two-component BECs around $\delta=1$. 
Using the imaginary time propagation of the time-dependent version of the 
Eqs.~\eqref{bindimless1} and \eqref{bindimless2}, we calculate the equilibrium solutions. 
We confirm the sufficient 
convergence of certain quantities such as the total energy of the system.
We conducted the simulations for several values 
of $\tilde{\Omega}$, $\delta$, and initial trial functions with the Gaussian form. 
The typical examples of the obtained structures are shown in Fig.~\ref{saka1}. 
As mentioned, the clear square lattice [Fig.~\ref{saka1}(a)] and vortex sheet structure [Fig.~\ref{saka1}(c)] 
appear for $\delta \lesssim 1$ and $\delta \gtrsim 1$, respectively. 
For $\delta=1$, the vortex lattice exhibits more complicated form as shown in Fig.~\ref{saka1}(b).
For the $\Psi_2$-component, a clear structure of the honeycomb structure 
can be seen. For the $\Psi_1$-component, the two vortex cores are closely approached 
at the cells of the honeycomb created by the $\Psi_2$-vortices. 
Thus, this structure has been referred to as a ``double-core lattice" \cite{Kasamatsu}. 
This indicates that the symmetry of the solutions is spontaneously broken 
because the inter-distance of the vortex core in each component is different, 
even for our symmetric parameter setting. 

\begin{figure}[ht]
\centering
\includegraphics[width=1.0\linewidth, bb=0 0 425 454]{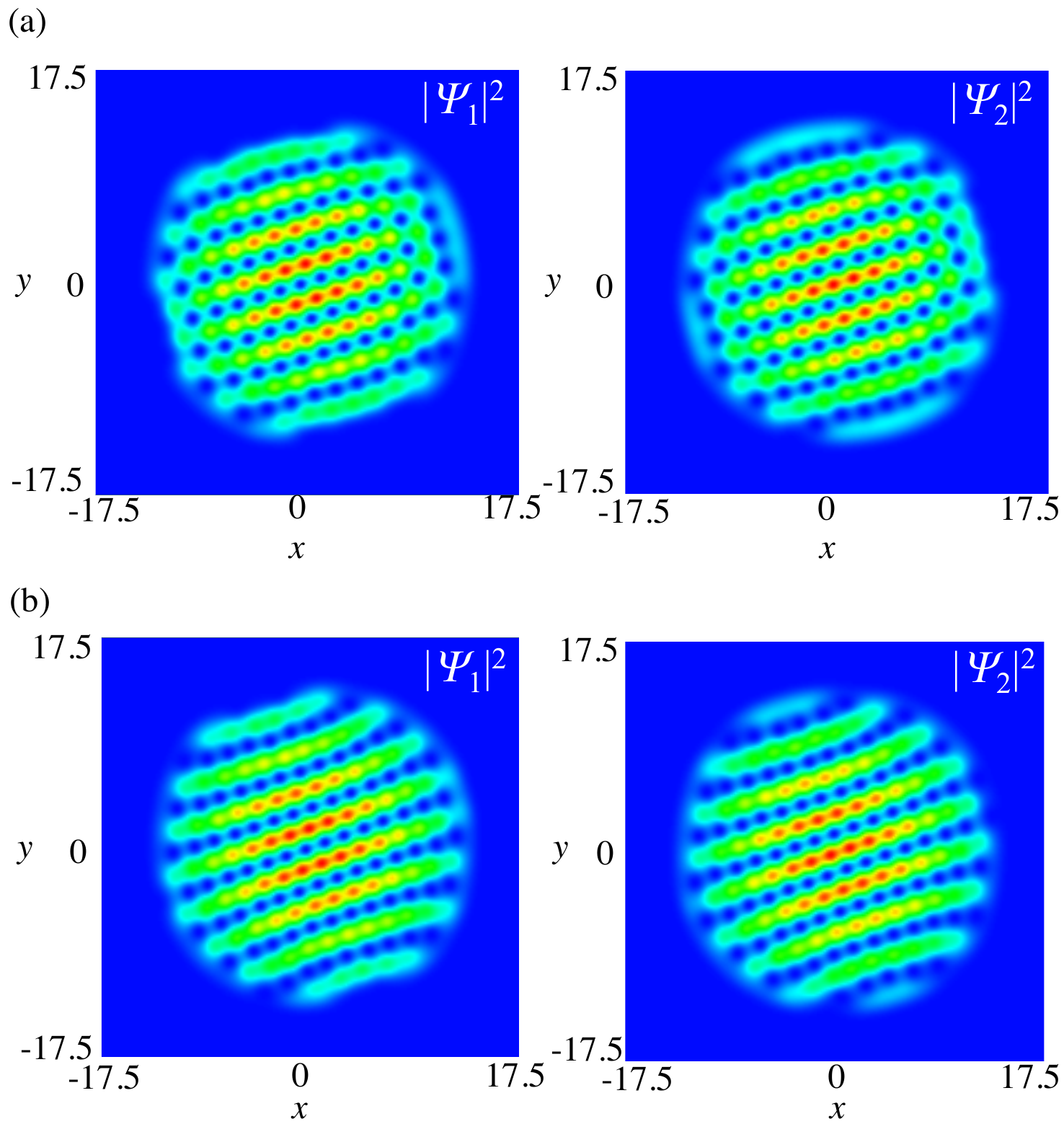}
\caption{(Color online) The density profile $|\Psi_1|^2$ (left) and $|\Psi_2|^2$ (right) of the 
equilibrium solutions of Eqs.~\eqref{bindimless1} and \eqref{bindimless2} 
for $C=4000$, $\tilde{\Omega}=0.8$. (a) for $\delta = 0.995$ and (b) for $\delta = 1.005$. The initial 
condition of the imaginary time evolution corresponds to the solution of Fig.~\ref{saka1}(b).}
\label{saka2}
\end{figure}

Throughout the numerical simulations starting from various initial conditions, we can 
observe various metastable configurations of the vortex lattices in addition to Fig.~\ref{saka1}(b). 
The typical one is a stripe structure, where vortices in each component align rows alternately 
(like those shown in Fig.~\ref{saka2}). 
The vortex stripe structure was predicted by Mueller and Ho as the 
stable lattice state for $\delta =1$ \cite{Mueller}. 
However, the perfect periodicity of the ansatz cannot describe the 
honeycomb lattice characterized by different periodicities as shown in Fig.~\ref{saka1}(b). 
In other cases, some defects remain in the honeycomb lattice, 
which decays very slowly during the imaginary time propagation. 
For example, there appears a domain wall separating two 
configurations with a honeycomb lattice and a double-core lattice 
with different inter-vortex distances. 

Next, in order to observe the stability of the honeycomb--double-core lattice, we start the imaginary time 
propagation from the solution of Fig.~\ref{saka1}(b) and 
change $\delta$ slightly from unity. We find that the honeycomb--double-core lattice is deformed to the stripe structure, 
as shown in Fig.~\ref{saka2}, for $|\delta -1| \geq 0.005$. This result indicates that the lattice structure is 
consistent with the prediction of Ref.~\cite{Mueller} except for $\delta =1$.
Therefore, the appearance of the honeycomb--double-core lattice is peculiar to the solution only for $\delta=1$.

\section{vortex lattice structure through SU(2) transformation}
According to the numerical simulation, the vortex lattice structure is 
anomalous only for $\delta = 1$, where the coupling constants are 
equivalent as $g_{1} = g_2 = g_{12}$. 
Then, the two component system in our case has an exact SU(2) symmetry, 
This is because the interaction energy of Eq.~\eqref{interactionene} can be simply written as: 
$E_\text{int} = (g/2) \int d \mathbf{r}  n_\text{T}^2$ with the total density 
$n_\text{T} = |\Psi_1|^2 + |\Psi_2|^2$. 
Then, the energy is invariant under the global SU(2) symmetric operation 
\begin{align}
\text{SU(2)} &= e^{-i \gamma \hat{\sigma}_z / 2}  e^{-i \beta \hat{\sigma}_y / 2}  e^{-i \alpha \hat{\sigma}_z / 2} \nonumber \\
&= \left( 
\begin{array}{cc}
\cos(\beta/2) e^{-i(\alpha+\gamma)/2} & -\sin(\beta/2) e^{i(\alpha-\gamma)/2} \\
\sin(\beta/2) e^{-i(\alpha-\gamma)/2} & \cos(\beta/2) e^{i(\alpha+\gamma)/2}
\end{array}
\right)
\end{align}
where $\bm{\sigma} = (\sigma_x,\sigma_y,\sigma_z)$ is the Pauli matrix 
and $(\alpha, \beta, \gamma)$ are the Euler angles. 
Here, we see that how the vortex lattice changes through the SU(2) transformation. 
Given that the rotation of the angles $\alpha$ and $\gamma$ only changes the global phase 
of the wave function by a constant, this is not relevant to the lattice structure. 
Thus, the structural change of the vortex lattice may occur through the variation of $\beta$, 
which represents the rotation of the spin space around the $y$-axis. 

\begin{figure}[ht]
\centering
\includegraphics[width=0.80\linewidth, bb=0 0 226 510]{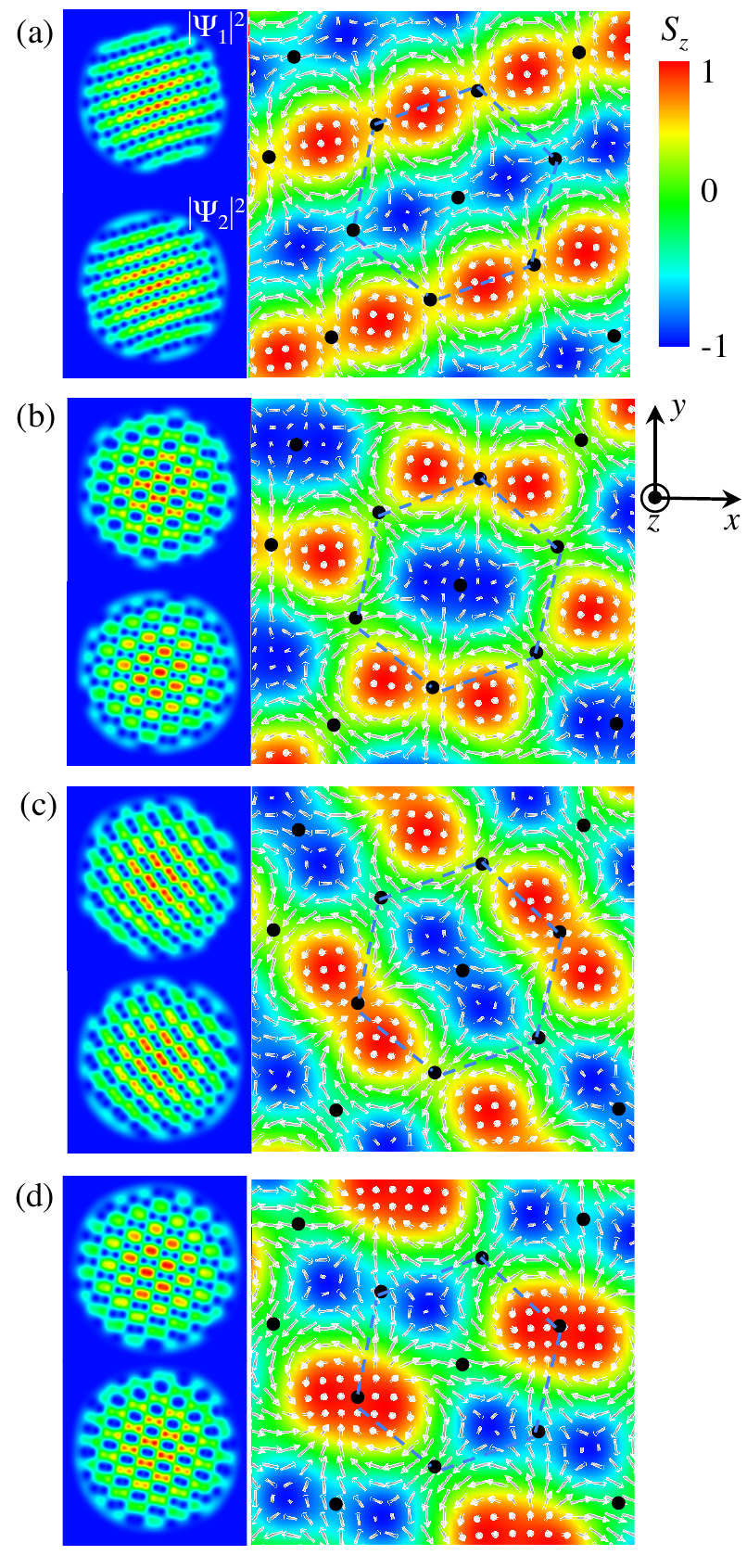}
\caption{(Color online) The left panels show the density profile $|\Psi_1|^2$ (top) and $|\Psi_2|^2$ (bottom) with 
the vortex lattice for $\delta=1$ through the change of the Euler angle $\beta$ from that of Fig.~\ref{saka1}(b). The values 
of $\beta$ are (a) $\pi/4$, (b) $\pi/2$, (c) $3\pi/4$, and (d) $\pi$. 
The profiles are plotted in the region $-17.5 \leq x,y \leq 17.5$. 
The right panels show the distribution of the spin $\bm{S}(\bm{r})$ of the corresponding solution 
in the region $-2.5 \leq x,y \leq 2.5$. The magnitude of $S_z$ is shown by the color scale. 
The black dots represent the 
points where the topological charge $q(\bm{r})$ vanishes; see Fig.~\ref{spin}(b).}
\label{su2}
\end{figure}
The left panels of Fig.~\ref{su2} show the change of the lattice structure through the rotation of $\beta$. 
Here, we start from the solution of Fig.~\ref{saka1}(b) as the state of $\beta=0$. 
We can see that the stripe state and the honeycomb--double-core lattice are connected 
\textit{continuously} by the $\pi/4$ rotation of $\beta$ as shown in Fig.~\ref{su2}(a). 
For $\beta=\pi/2$ the structure is similar to that in Fig.~\ref{saka1}(b), but the pairs of the vortices forming double cores 
are exchanged with the nearest neighbor ones.  
For $\beta=3\pi/4$ the condensates form a nested structure consisting of lattices of vortex pairs 
as seen in Fig.~\ref{su2}(c), 
where the polarization of the pairs is almost perpendicular to the stripes at $\beta=\pi/4$. 
With further increasing $\beta$ to $\pi$, the structure of $\Psi_1$- and $\Psi_2$-component 
is interchanged with that in Fig.~\ref{saka1}(b) [Fig.~\ref{su2}(d)]. 
It is important to mention that these structures are degenerate with respect to the 
SU(2) transformation. 
Hence, one trial of the imaginary time propagation of the GP equation selects one of 
the degenerate states. This is the reason why the numerical calculations yield 
lots of metastable configurations. 
 
It is fruitful to see this structural transformation from the viewpoint of the pseudospin texture. 
The two-component BECs can be represented by a two-component spinor and 
the local pseudospin of the two-component system can be defined by 
$\bm{S} = \bm{\Psi}^{\dagger} \bm{\sigma} \bm{\Psi} / n_\text{T}$ 
with $\bm{\Psi} = (\Psi_1, \Psi_2)^\text{T}$ \cite{Kasarev2}. 
The right panel of Fig.~\ref{su2} shows the pseudospin profile of the corresponding state for each $\beta$. 
The spin points out of (into) the page at the vortex center of $\Psi_1$ ($\Psi_2$), around which the spin direction 
covers the north hemisphere (south hemisphere) of the spin space, e.g., $2\pi$ steradian of the solid angle. 
This structure is known as the half-skyrmion (meron). 
For the vortex stripe ($\beta=\pi/4$), the pseudospin forms an alternative array 
of the half-skyrmions with the spins at the center point up and down. 
Let us see the structure in the hexagonal cell shown by the dashed line in Fig.~\ref{su2}(a).  
There are four cores of the skyrmions where the spins are pointing into or out of the page. 
By globally rotating the spins by $\pi/4$ around the $y$-axis, there appears a region with 
$S_z \sim 1$ at the center of the hexagon, around which the directions of the spins cover 
the south hemisphere twice [Fig.~\ref{su2}(b)]. Thus, this structure corresponds to the doubly-winding 
half-skyrmion. Mueller showed that the doubly-winding half-skyrmion and the 
four half-skyrmions are connected through the rotation of the angle $\beta$ \cite{Muellerspin1}. 
This doubly-winding half-skyrmion is surrounded by the six half-skyrmions. 
These composites can be seen as the honeycomb and double-core lattices in the density profile.  
With increasing $\beta$ further, the doubly-winding half-skyrmion again splits into two half-skyrmions into 
the direction perpendicular to the original stripe [Fig.~\ref{su2}(c)]. For $\beta=\pi$ [Fig.~\ref{su2}(d)] the doubly-winding 
half-skyrmion with $S_z=+1$ at the core appears next to the noticing hexagon. 

\begin{figure}[ht]
\centering
\includegraphics[width=1.0\linewidth, bb=0 0 822 227]{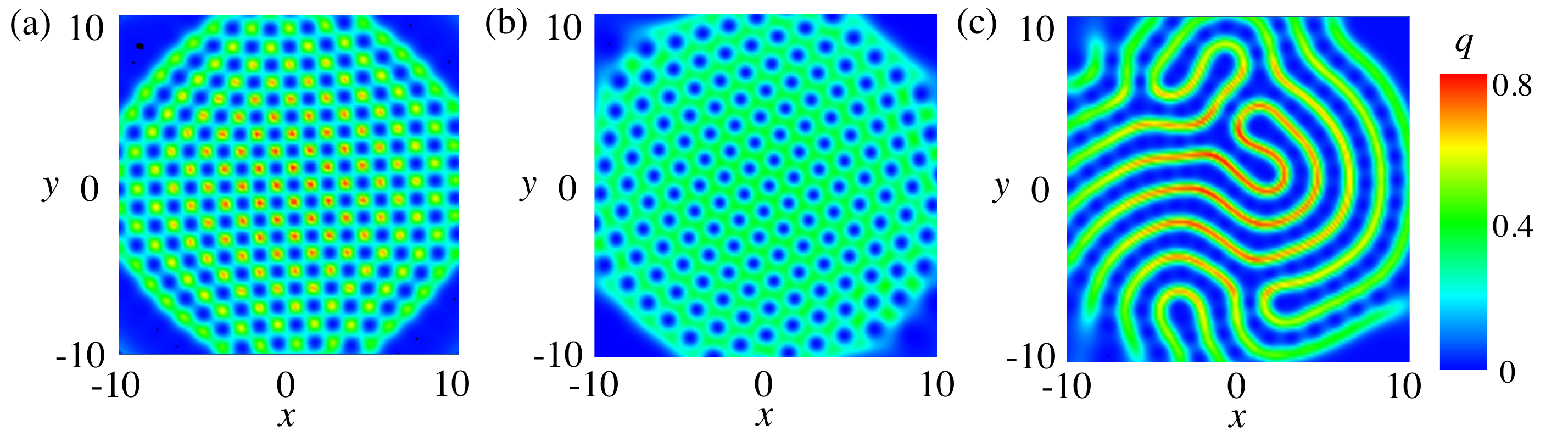}
\caption{(Color online) The profile of the topological charge density for 
$\Omega=0.8$ and (a) $\delta=0.9$, (b) $\delta=1.0$, and (c) $\delta=1.1$, 
corresponding to the solution of Fig.~\ref{saka1}. 
The topological charge is calculated within the Thomas-Fermi radius.}
\label{spin}
\end{figure}
We also calculate the profile of the topological charge density defined by 
\begin{equation}
q(\bm{r}) = \frac{1}{4\pi} \bm{S} \cdot \left( \frac{\partial \bm{S}}{\partial x} \times \frac{\partial \bm{S}}{\partial y} \right)
\end{equation}
In Fig.~\ref{spin}, we plot the profile of the topological charge corresponding to the solutions of Fig.~\ref{saka1}.
We can see the clear structural change between the typical lattice structures. 
The stripe and honeycomb--double-core lattice corresponds to the hexagonal lattice of the 
``holes" of the topological charge $q(\bm{r})$ as shown in Fig.~\ref{spin}(b), which is clearly distinguished from the square 
lattice for $\delta \lesssim 1$ [Fig.~\ref{spin}(a)] and the vortex sheets for $\delta \gtrsim 1$ [Fig.~\ref{spin}(c)]. 
The hole of $q(\bm{r})$ is a clear signature of the doubly-winding half-skyrmion \cite{Muellerspin1}, 
where the topological charge is distributed around the skyrmion core. The magnitude of $q(\bm{r})$ in Fig.~\ref{spin}(b) 
is almost half of that in (a) and (c).

Note that an exact SU(2) symmetry operation is impossible in our case, 
because each particle number of the two-component BEC must be conserved. 
The change of $\beta$ causes a population transfer between the two components. 
However, the approximate SU(2) operation is possible because of the 
presence of the vortices, which disturbs the spatial phase distribution of 
the condensate wave function. 
This can be seen by the fact that the norm of the wave function after the SU(2) 
transformation $\bm{\Psi} \to \bm{\Psi}'$ can be 
written as: 
\begin{align}
\int d\mathbf{r} |\Psi_i'|^2 = & \int d\mathbf{r} \biggl[ \cos^2 \frac{\beta}{2} |\Psi_i|^2 + \sin^2 \frac{\beta}{2} |\Psi_j|^2  \nonumber \\
& \mp 2 \cos \frac{\beta}{2} \sin \frac{\beta}{2} |\Psi_i| |\Psi_j| \cos (\theta_2 - \theta_1 -\alpha) \biggr].
\end{align}
If the third term of the right-hand side vanishes, the SU(2) operation does not change 
each of the particle numbers. 
Now the relative phase $\theta_2-\theta_1$ is not physically relevant in our case, because 
there is not Josephson-like coupling between the two components \cite{Cipriani,Uranga}. 
However, the presence of vortices yields the periodic spatial variation of $\theta_2 -\theta_1$, 
such that the spatial integral of the third term can vanish approximately. 
Thus, the spin rotation of the angle $\beta$ can be effectively achieved. 
This is the reason why the vortex lattice structure is fragile in this SU(2) symmetrical case. 
The presence of many vortices opens the door for the structural change associated 
with the SU(2) degeneracy. 

\section{Summary and Discussion}
In this paper, we discuss the vortex lattice structure in rotating two-component 
BECs with equal intra- and intercomponent interaction strength. 
Then, the hamiltonian is rendered invariant under the SU(2) operation. 
The resulting degeneracy brings forth a rich variety of configurations of the 
vortex lattices, where the honeycomb double-core lattices and stripes can 
coexist. These structures can be connected by a continuous global SU(2) 
transformation. 
Thus, the prediction by Mueller and Ho \cite{Mueller} is valid even at this SU(2) symmetric 
point. 

In the experiments, the approximate SU(2) symmetry holds for the two-component 
BECs of $^{87}$Rb with the hyperfine spins $| 1, -1\rangle$ and $| 2, 1\rangle$. 
In the experiments, the authors observed that the vortex lattices are fragile 
in the early stages of the dynamics before demonstrating the order 
of the square structures. 
Although this represents the nonequilibrium situation, it might be partly 
due to the degeneracy of the system. 
It is interesting to observe the stripe or honeycomb--double-core structures by tuning 
the coupling strengths to satisfy the SU(2) symmetry exactly. 
In other words, the observation of the fragile structure could be a consequence of the 
SU(2) symmetry. 
In order to see a rigid honeycomb--double-core 
structure, further fine tuning of the parameters, such as the $s$-wave scattering 
lengths \cite{Thalhammer,Papp}, is necessary to realize the scenario. 
In addition, our argument would be useful for discussing the vortex lattices in spinor BECs 
characterized by high symmetry groups. 

\acknowledgements
This work was partly supported by KAKENHI from JSPS (Grant No. 26400371).

\end{document}